\documentclass[authoryear,preprint,review,11pt]{elsarticle}

\usepackage{amssymb}

\usepackage{lineno}
\usepackage{hyperref}
\usepackage[utf8]{inputenc}
\usepackage{multirow}
\usepackage{url}
\usepackage[none]{hyphenat}
\usepackage[flushleft]{threeparttable}
\usepackage{textcomp}

\journal{Journal of Systems and Software}

\begin{document}

\begin{frontmatter}



\title{Descriptions of issues and comments for predicting issue success in software projects}


\author{Sandra L. Ramírez-Mora\corref{cor1}\fnref{label1}}


\ead{sandra.ramirez@ciencias.unam.mx}
\cortext[cor1]{Corresponding author}

\author[label2]{Hanna Oktaba}
\author[label3]{Helena Gómez-Adorno}
\address[label1]{Posgrado en Ciencia e Ingeniería de la Computación, Universidad Nacional Autónoma de México, Mexico City, Mexico}
\address[label2]{Facultad de Ciencias, Universidad Nacional Autónoma de México, Mexico City, Mexico }
\address[label3]{Instituto de Investigaciones en Matemáticas Aplicadas y en Sistemas, Universidad Nacional Autónoma de México, Mexico City, Mexico}

\begin{abstract}
\small
Software development tasks must be performed successfully to achieve software quality and customer satisfaction. Knowing whether software tasks are likely to fail is essential to ensure the success of software projects. Issue Tracking Systems store information of software tasks (issues) and comments, which can be useful to predict issue success; however; almost no research on this topic exists.
This work studies the usefulness of textual descriptions of issues and comments for predicting whether issues will be resolved successfully or not.
Issues and comments of 588 software projects were extracted from four popular Issue Tracking Systems. Seven machine learning classifiers were trained on 30k issues and more than 120k comments, and more than 6000 experiments were performed to predict the success of three types of issues: bugs, improvements and new features.
The results provided evidence that descriptions of issues and comments are useful for predicting issue success with more than 85\% of accuracy and precision, and that the predictions of issue success vary over time. Words related to software development were particularly relevant for predicting issue success.
Other communication aspects and their relationship to the success of software projects must be researched in detail using data from software tools.

\end{abstract}






\end{frontmatter}



\small
© 2020. This manuscript version is made available under the CC-BY-NC-ND 4.0 license http://creativecommons.org/licenses/by-nc-nd/4.0/

The final version is available at: \url{https://doi.org/10.1016/j.jss.2020.110663}

\normalsize

\section{Introduction}
\label{S:1}

\sloppy

Software activities, such as the development of new functionalities, the improvement of software, and the correction of defects must be continuously and successfully performed to ensure software quality and to satisfy the needs of customers and users of software products. Predicting whether software activities will be performed successfully is crucial in software projects, especially when a lot of people use software products that constantly evolve. 
Knowing whether an activity is likely to fail can increase software quality and customer satisfaction in software projects because resources can be managed to ensure the completion of software activities, such as the correction of blocking software defects that are crucial for many stakeholders of software products. Early prediction of task success can also reduce development time because people can make opportune decisions without having to wait until the completion of an activity.

Research on predicting outcomes of software projects often focus on predicting the time or effort to perform software tasks; however, knowing the approximate time in which a software task will be completed may be useless if the task will be unsuccessfully completed and actions for ensuring the success of the task are not performed, so the prediction of issue success is crucial in software engineering.

Software development tools facilitate collaboration, are useful for managing software activities, and store information that can be useful to predict the success or failure of software tasks. Particularly, Jira is a popular software development tool in which software tasks (called “issues”) can be recorded and managed. Comments of issues can also be recorded in Jira to provide additional detail about software tasks and collaborate with team members. Jira is often used as an Issue Tracking System (ITS) and many organizations such as Apache and Spring have a public Jira ITS. Comments of issues that are recorded in Jira ITSs usually contain implicit and explicit information about the progress and development of software tasks (issues), so they can be used to detect whether an issue will be successfully resolved or not.

Software teams frequently use web tools (such as ITSs) to communicate; therefore, textual descriptions of comments and issues from Jira ITSs represent an important part of the information that is communicated among customers, users, and developers of software products. Communication is closely related to the success of software projects, and some works have found a positive impact of quality communication on productivity and project success \citep{DEOMELO2013, Destefanis2016, SMS2017}. Based on the relationship between communication and the success of software projects, and based on the fact that comments reflect the nature of communication in software development and contain implicit and explicit information about the progress of issue resolution, the following hypothesis was formulated: descriptions of issues and comments from Jira ITSs are useful for predicting issue success.
From this hypothesis, the following questions were defined to study the usefulness of descriptions of issues and comments for predicting the success of three types of issues (bugs, improvements and new features).

RQ$_1$: Are textual descriptions of issues and comments from Jira ITSs useful to predict issue success in software projects?

RQ$_{1.1}$: What kind of information is useful to predict issue success?

RQ$_{1.2}$: How does the prediction of issue success vary over time?

RQ$_{1.3}$: How does the prediction of issue success vary with respect to bugs, improvements and new features?
\hfill \break
This work contributes on the research regarding the early prediction of issue success using descriptions of comments and issues from Jira ITSs. The remainder of this paper is organized as follows: section \ref{S:2} describes the research background; section \ref{S:3} describes the related work; section \ref{S:4} details the research method; section \ref{S:5} describes the results; in section \ref{S:6}, the results and validity of this work are discussed; and in section \ref{S:7}, conclusions are described and directions for future work are suggested.

\section{Background}
\label{S:2}

Software development tools are very useful for communicating and collaborating during software development processes, especially for people that are geographically distributed. During software lifecycle, developers, users, and customers require to report software defects, provide feedback, request new features or improvements, and monitor the progress of software tasks, and software development tools are very useful for these purposes. Knowing the progress of software tasks is very important for users and customers of software products because they need to know in advance whether their requests are being successfully achieved.

Jira is a software development tool widely used by agile teams to plan, track, and release software. In Jira, ``issues'' are the elemental components of a software project and represent software bugs, project tasks, requirements, improvements or another issue type. Comments of issues can also be recorded in Jira to provide additional information of issues and facilitate the collaboration among people (developers, users, customers) involved in software development activities. Jira is commonly used as an Issue Tracking System (ITS) and many organizations  (including Apache, Spring, Hibernate, Atlassian, Red Hat and Fedora) have a public Jira ITS to record and manage issues of their software projects, particularly of their open source projects.

\begin{figure}[hbt!]
\centering\includegraphics[width=0.5\linewidth]{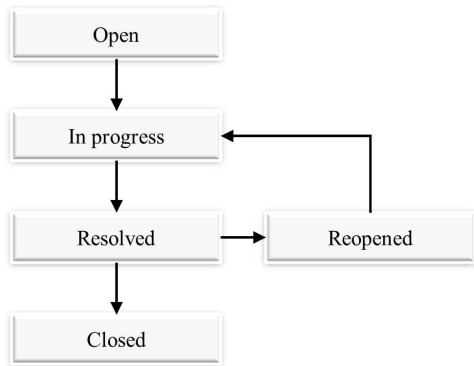}
\caption{Example of some possible statuses of an issue during its lifecycle}
\label{fig:lifecycle}
\end{figure}

Progress of tasks can be monitored in Jira because issues are continually tagged according to their status during their lifecycle. An issue is initially ``open'' (created), and when people start working on it, the issue is tagged as “in progress”. When an issue is attended, it is tagged as “resolved”, and if the resolution is accepted, the issue is tagged as “closed”, otherwise, the issue is “reopened”. Authorized stakeholders (developers, project managers, business people, etc.) in a software project can open, resolve, and close issues, and they can indicate issue labels, such as the status and resolution of issues. Figure \ref{fig:lifecycle} shows an example of some possible statuses of an issue during its lifecycle.

\begin{figure}[hbt!]
\centering\includegraphics[width=0.7\linewidth]{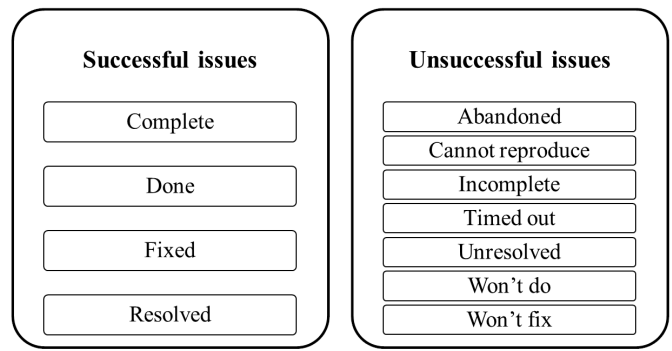}
\caption{Usual resolution tags of successful and unsuccessful issues}
\label{fig:success}
\end{figure}

When an issue is resolved, it is tagged according to its resolution type (Figure \ref{fig:success}). Issues that have been resolved successfully are usually tagged as “complete”, “done”, “fixed or “resolved” in Jira ITSs; however, there are many issues that are not resolved successfully, and they are typically tagged with one of the following labels.

\begin{itemize}
\item “Abandoned”. Indicates that an issue was abandoned.
\item “Cannot reproduce”. Indicates that an issue cannot be reproduced.
\item “Incomplete”. Indicates that an issue is not described completely.
\item “Timed out”. Indicates that an issue was closed due to lack of response.
\item “Unresolved”. Indicates that an issue was not resolved.
\item “Won’t do”. Indicates that an issue won't be actioned.
\item “Won't fix”. Indicates that the problem described is an issue which will never be fixed.
\end{itemize}

In Jira ITSs, each issue and comment is stored with a description, reporter, a unique identifier and the date of report. Descriptions of comments and issues are texts that have lexical, syntactic and semantic characteristics, so they can be studied using Natural Language Processing (NLP) techniques. Descriptions of comments and issues often include technical information (URLs, fragments of code, software specifications, file directories), information about the progress of software tasks, and many of them include information regarding interactions and collaboration among people. 

Comments and issues from Jira ITSs represent part of the communication that is performed during software development activities. Some communication models represent communication as an action in which a sender (source or speaker) transmits a message though some channel to a receiver~\citep{encyclopedia2009}. Considering this kind of models, the reporter of a comment or issue is the sender, the description of the comment or issue is the message, and the people who read the comment or issue are the receivers in a communication process. Comments and issues are usually reported and read by developers, project managers, customers and users of software products.

Communication is a key factor for the success of software projects and the productivity improvement, and some previous works provided evidence of this. In a systematic mapping study~\citep{SMS2017}, communication was found to be a factor that impacts productivity in agile software development. \cite{FAGERHOL2015} concluded that enhancing performance experiences of software teams requires integration of soft factors, such as communication. \cite{Hoegl2001} and \cite{LINDSJORN2016} found a strong correlation between teamwork quality (which includes communication) and the success of software development teams. \cite{Nasir2011} conducted a comparative study and found that communication is one of the critical success factors for software projects.
The results of \cite{Destefanis2016} showed that the level of politeness in the communication process among developers does have an effect on the
time required to fix issues in software projects. \cite{McLeod2011} conducted a survey of research regarding the factors that affect software systems development project outcomes, and found that communication is often perceived as an important dimension of the interaction between users and development staff, essential for effective functioning of the project team, and a key factor in system success; these declarations are based on the works of \cite{Akkermans2002}, \cite{Butler2003}, \cite{Butler2001}, \cite{Hartwick2001}, \cite{Sawyer1998}, and \cite{Somers2001}. \cite{Garousi2019} studied the correlation of critical success factors with the success of software projects and found that the higher the quality of internal team communication, the higher the team building and team dynamics at the end of a project.

Based on the evidence of the close relationship between communication and the success of software projects, and the kind of information that is usually included in comments and issues (such as the progress of software tasks) from ITSs, the prediction of issue success could be performed using the textual descriptions of issues and comments, which are communicated during software development activities. The prediction of issue success is a concern of stakeholders (developers, customers, users) of software products, who frequently need to know whether an issue will be successfully addressed or not; however, scarce research on this topic exists.

\section{Related work}
\label{S:3}
Researchers and practitioners are often interested on predicting and studying outcomes of software processes and tasks (such as time, effort, cost and success) and characteristics of software products (such as quality, faults and bugs). \cite{Murguia} investigated the influence of the maintenance activities types on issue resolution time using data from projects in GitHub.

Machine learning techniques are usually used to predict software development aspects. \cite{Catal2011} conducted a literature review of the trends on software fault prediction and found that supervised algorithms of machine learning and software metrics can be used to build prediction models. \cite{Hall2011} conducted a systematic literature review to investigate how the context, variables and modeling techniques influence the performance of fault prediction models. They found that many different variables have been used (including metrics of process and products, metrics relating to developers, and texts of source code) and concluded that more studies with a reliable methodology and detailed content are needed.

Most of the works that have reported predictions in software development have used data of software artifacts and processes, such as the work of \cite{Guo2010}. \cite{Suma2014} used a machine learning classifier to predict software defects using data of software defects, software size and project development time. In the work of \cite{Choetkiertikul2018}, an approach was proposed for predicting delivery capacity for software development iterations applying machine learning techniques and using data from previous iterations and issues. 

Few works that have used data of human factors for predicting outcomes of software tasks exist. \cite{DucAnh2011} assessed different types of issue lead time prediction models using human factor measures that were collected from ITSs. These data include the experience of reporters and assignees of issues, the number of comments of issues, and the number of the involved stakeholders in an issue. The results indicated that the number of stakeholders and the average of the lead time of previous issues (resolved by developers) are important variables in constructing prediction models of issue lead time; however, authors concluded that more variables should be explored to achieve better prediction of performance. \cite{Ortu2015} studied the relationship between “affectiveness” (sentiments, emotions and politeness) of developers and the time to fix issues using machine learning techniques. The authors used data of issues and comments and found that happy developers are likely to fix issues in short time and that negative emotions are linked with longer issue fixing time. 

Some works that report the use of texts to perform predictions of software tasks outcomes and software products characteristics exist; however, scarce research regarding the prediction of software tasks success exists.  \cite{DiSorbo2019} investigated the nature of “won't fix” issues in GitHub, performed predictions of “won't fix” issues using machine learning techniques and textual features (titles and descriptions of reported issues), and used such textual features to identify the most important words in the issue titles and descriptions.  \cite{FRONZA2013}  described an approach to predict failures of software systems based on log files using text analysis techniques and machine learning algorithms. \cite{BINKLEY2009} applied three language-processing measures (based on the percentage of natural language words in code, the percentage of identifiers that violate syntactic conciseness and consistency rules, and the similarity between a module’s comments and its code) to the problem of fault prediction using data from Mozilla and their results demonstrated the usefulness of these measures for fault prediction. \cite{VALDIVIAGARCIA2018} studied bugs that block the fixing of other bugs in eight open source projects and proposed a model to predict them using data of bugs such as priority, severity, descriptions and comments. They found that the description and the comments included in the bugs were the most important factors for predicting blocking bugs.

In general, almost no works on predicting issue success exists, and few works have reported the use of human factors and interaction aspects (such as natural language and communication) to perform predictions in software development. Almost no works that have used textual descriptions (and its lexical, syntactic and semantic characteristics) to perform success predictions exist. In addition, scarce research on the identification of relevant information from texts to predict issue success exists, and works that consider issue types and time as variables for the prediction of issue success are lacking. The objective of this work is to contribute to the investigation on these topics.

\section{Research method}
\label{S:4}

This work addresses the following questions.

RQ$_1$: Are textual descriptions of issues and comments from Jira ITSs useful to predict issue success in software projects?

RQ$_{1.1}$: What kind of information is useful to predict issue success?

RQ$_{1.2}$: How does the prediction of issue success vary over time?

RQ$_{1.3}$: How does the prediction of issue success vary with respect to bugs, improvements and new features?
\hfill \break
The research questions can be answered using a machine learning approach. From the machine-learning perspective, the issue success prediction can be viewed as a binary-class classification problem, in which automatic methods have to assign positive or negative class labels (success or not success) to objects (texts).

\begin{figure}[hbt!]
\centering\includegraphics[width=1.0\linewidth]{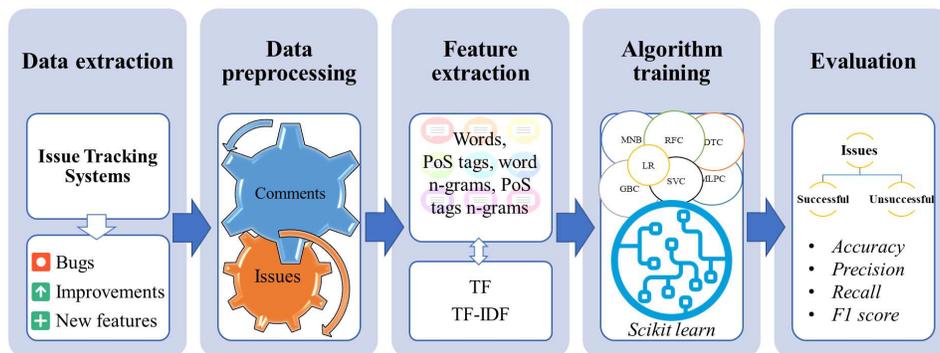}
\caption{Machine learning approach}
\label{fig:method}
\end{figure}

In order to train a model that is able to perform predictions of issue success, a dataset for the learning process is required. With this aim, a set of comments and issues from ITSs were first collected.
Texts were preprocessed to be easily understood by machine learning algorithms and to delete irrelevant information from them. Relevant features (characteristics) were extracted from the preprocessed texts, and the texts were transformed into a numerical representation in the form of a vector; then, machine learning algorithms were trained on the vectorized texts to produce a prediction model, which was used to predict issue success. The general steps of the machine learning approach are shown in Figure \ref{fig:method} and are detailed in the following sections.

\subsection{Data extraction}

The data extraction process was performed to collect the data that were required to train classification algorithms in order to perform predictions of issue success and answer the research questions. 
The following four public Jira ITSs store a considerable number of software projects (many of them open source) that are of the interest of many people dedicated to software engineering, so they were selected as data source.

\begin{itemize}
\item Apache’s JIRA Issue Tracking System\footnote{Apache’s ITS: https://issues.apache.org/jira/secure/Dashboard.jspa} is an open Issue Tracking System that stores more than 600 software projects.
\item Atlassian’s public Issue Tracking System\footnote{Atlassian’s ITS: https://jira.atlassian.com/secure/BrowseProjects.jspa} is used to manage more than 30 software projects of Confluence, Bitbucket, Jira and other Atlassian products.
\item Red Hat’s Issue Tracking System\footnote{Red Hats’s ITS: https://issues.redhat.com/secure/Dashboard.jspa} stores more than 400 software projects including Red Hat projects. 
\item Spring’s Issue Tracking System\footnote{Spring’s ITS: https://jira.spring.io/secure/Dashboard.jspa} is a system for tracking issues, progress, and roadmaps for more than 80 Spring projects and their derivatives.
\end{itemize}

 With the aim of studying the success of different types of software tasks, the following three types of issues were extracted from the above Jira ITSs using the Jira API \footnote{JIRA Agile REST API Reference: https://docs.atlassian.com/jira-software/REST/7.0.4/} and the web pages of the Jira ITSs.

\begin{itemize}
\item	Bugs. Software defects and failures that affect software outcomes.
\item	Improvements. Software upgrades and enhancements.
\item	New features. New functionalities of software systems.
\end{itemize}

In order to study a sufficient number of issues of each type (at least 10K), issues of 588 software projects (333 from the Apache’s ITS, 14 from the Atlassian’s ITS, 191 from the Red Hat’s ITS, and 50 from the Spring’s ITS) were extracted. These projects represent about 50\% of the projects stored in the mentioned repositories and were selected to study different software systems such as frameworks, software extensions, servers, libraries and web components. The selected projects vary on size and the number of people involved on them. Table \ref{table:Projects1} shows the distribution of projects according to their size (number of issues), and Table \ref{table:Projects2} shows the distribution of the projects according to their number of developers (assignees), commenters and watchers of the studied issues. Most of the projects are developed using Java and JavaScript as programming languages.

\begin{table}[hbt!]
\small
\begin{threeparttable}[t]
  \begin{tabular}{ |p{2.2cm}||p{2.2cm}p{2.2cm}p{2.2cm}p{2.8cm}|}
  \hline
  \multicolumn{5}{|c|}{\textbf{\textit{Projects by total number of issues}}}\\
  \hline
    \multirow{2}{*}{Repository}&
  \multicolumn{4}{c|}{\textit{Number of issues}}\\
 
\cline{2-5} &
0 - 1000&1001 - 2000&2001 - 3000&More than 3000\\
\hline

Apache&204&40&29&60\\
Atlassian&3&2&0&9\\
Red Hat&123&25&15&28\\
Spring&35&9&1&5\\
\hline

  \hline
  \multicolumn{5}{|c|}{\textbf{\textit{Projects by number of studied issues\tnote{a}}}}\\
  \hline
    \multirow{2}{*}{Repository}&
  \multicolumn{4}{c|}{\textit{Number of issues}}\\
 
\cline{2-5} &
0 - 500&501 - 1000&1001 - 1500&More than 1500\\
\hline

Apache&222&39&25&47\\
Atlassian&5&1&4&4\\
Red Hat&147&20&6&18\\
Spring&38&8&1&3\\
\hline

 \hline
 \end{tabular}

 \begin{tablenotes}
     \item[a]Closed bugs, new features and improvements.
   \end{tablenotes}
   
\caption{Distribution of the studied projects by number of issues}  
\label{table:Projects1}
 \end{threeparttable}%

\end{table}

\begin{table}[hbt!]
\small
  \begin{tabular}{ |p{2.2cm}||p{2.2cm}p{2.2cm}p{2.2cm}p{2.8cm}|}
 
 \hline
  \multicolumn{5}{|c|}{\textbf{\textit{Projects by number of assignees}}}\\
  \hline
    \multirow{2}{*}{Repository}&
  \multicolumn{4}{c|}{\textit{Number of assignees}}\\
 
\cline{2-5} &
0 - 200&201 - 400&401 - 600&More than 600\\
\hline

Apache&227&35&18&53\\
Atlassian&13&1&0&0\\
Red Hat&191&0&0&0\\
Spring&50&0&0&0\\

\hline
 \hline
  \multicolumn{5}{|c|}{\textbf{\textit{Projects by number of commenters}}}\\
  \hline
    \multirow{2}{*}{Repository}&
  \multicolumn{4}{c|}{\textit{Number of commenters}}\\
 
\cline{2-5} &
0 - 200&201 - 400&401 - 600&More than 600\\
\hline

Apache&199&57&28&49\\
Atlassian&3&1&0&10\\
Red Hat&143&23&10&15\\
Spring&36&6&4&4\\

\hline

\hline
  \multicolumn{5}{|c|}{\textbf{\textit{Projects by number of watchers}}}\\
  \hline
    \multirow{2}{*}{Repository}&
  \multicolumn{4}{c|}{\textit{Number of watchers}}\\
 
\cline{2-5} &
0 - 1000&1001 - 2000&2001 - 3000&More than 3000\\
\hline

Apache&258&25&12&38\\
Atlassian&4&1&4&5\\
Red Hat&160&12&9&10\\
Spring&45&2&0&3\\

\hline
 \end{tabular}

\caption{Distribution of the studied projects by assignees, commenters and watchers considering the studied issues}  
\label{table:Projects2}
\end{table}

The issues of  the selected projects are publicly visible, are mostly written in English, and were registered from April 2001 to September 2019. The status of the extracted issues is “Closed”, which indicates that a resolution for each issue exists and that the issues are considered finished. The extracted issues are tagged according to their resolution (Figure \ref{fig:success}), so the issues that were resolved successfully and the issues that were resolved unsuccessfully were easily identified.
Data of issues (textual description, issue type, date of creation, date of resolution, date of last update, status and resolution) and data of comments (id, textual description, id of the issue they belong and date of reporting) were stored in a local database with the aim of facilitating the preprocessing process. The complete dataset is available online \citep{DataSet}.

Thirty thousand issues (15,000 successfully resolved and 15,000 unsuccessfully resolved) and their comments (about 120k) were particularly studied as a representative sample of the issues of the extracted projects. About 10,000 issues of each type (5,000 successfully resolved and 5,000 unsuccessfully resolved) were selected, so the dataset is balanced with respect to both, issue type and issue successfulness. The detailed distribution of the studied issues and comments is shown in Table \ref{table:generalData}. A balanced dataset was selected with the aim of studying a sufficient number of successful and unsuccessful issue types: the extracted software projects have, in general, more successful issues than unsuccessful issues (only 10\%-20\% of issues that are resolved in less than 30 days are unsuccessful) and most of the issues are bugs (improvements and new features represent a very small part of the total number of issues).

\begin{table}[hbt!]
\small
  \begin{tabular}{ |p{2.4cm}||p{2.2cm}p{2.2cm}||p{2.2cm}p{2.2cm}|}
  \hline
  
  \multirow{2}{*}{\textbf{Issue type}}&
  \multicolumn{2}{c||}{\textbf{Successful issues}}&
  \multicolumn{2}{c|}{\textbf{Unsuccessful issues}}\\
   \cline{2-5}
  & \textit{Issues} & \textit{Comments}& \textit{Issues} & \textit{Comments}\\
  \hline
Bugs&$\approx{5000}$& $\approx{25000}$ &$\approx{5000}$ & $\approx{17000}$\\
Improvements&$\approx{5000}$& $\approx{21000}$ &$\approx{5000}$& $\approx{19000}$\\
New features&$\approx{5000}$& $\approx{22000}$ &$\approx{5000}$& $\approx{19000}$\\

\hline
 \textbf{Total}&\textit{$\approx{15000}$}&\textit{$\approx{68000}$}&\textit{$\approx{15000}$}&\textit{$\approx{55000}$}\\
  \hline
  \end{tabular}

\caption{Distribution of the studied issues and comments}  
\label{table:generalData}
\end{table}

In addition to the selected issues in Table \ref{table:generalData}, the project with the greatest number of issues of each repository was selected to be studied in detail. This was performed with the aim of studying prediction of issue success in particular real projects. The selected projects are shown in Table \ref{table:projectData}.

\begin{table}[hbt!]
\small
\begin{threeparttable}[t]
  \begin{tabular}{ |p{2.4cm}||p{2.2cm}|p{2.2cm}|p{2.2cm}|p{2.2cm}|}
  \hline
  
 {} &\textit{Project A} & \textit{Project B}& \textit{Project C} & \textit{Project D}\\
  \hline
Key&FLEX&JSWSERVER&JBIDE&SPR\\
Repository&Apache&Atlassian&Red Hat&Spring\\
Issues (total) &35382&12559&26110&17413\\
Studied issues\tnote{a}&18209&3076&13122&8714\\
Comments\tnote{a}&55014&7210&72754&34925\\
Watchers\tnote{a}&490&11248&25239&15916\\
Assignees\tnote{a}&138&109&108&39\\
Commenters\tnote{a}&143&2194&1189&3136\\

  \hline
  \end{tabular}
  
\begin{tablenotes}
     \item[a]Closed bugs, new features and improvements.
   \end{tablenotes}
   
\caption{Projects selected to be studied in detail}  
\label{table:projectData}
 \end{threeparttable}%
\end{table}

\subsection{Preprocessing}

The following tasks were performed to eliminate irrelevant information from issues and comments after the extraction process. In addition, the preprocessing aims at reducing the amount of information to be processed, and improving prediction tasks.

\begin{itemize}
\item URLs were replaced with the string “url\_specification".
\item	Specific user names were identified and replaced with the string “user\_specification".
\item	Some comments included the string “+1", which means that the reporter of the comment expressed a positive vote to resolve the related issue, so this string was replaced with “vote\_specification".
\item	Numbers and software versions that were identified in the texts were replaced with “number\_specification" and “version\_specification" respectively.
\item	Specific emails were replaced with the string “email\_specification".
\item	Some comments included fragments of software code, which were replaced with the string “code\_specification".
\item	Some comments included paths to specify file directories, which were replaced with the string “path\_specification".
\end{itemize}

The preprocessed issues and their comments were tagged as “successful” or “unsuccessful” according to the resolution of each issue (Figure \ref{fig:success}). This tagging was required to perform supervised predictions using machine learning classifiers, which require a set of texts labeled with the class they belong.
Issue resolution time (in days) was calculated by subtracting the date of creation to the date of resolution of each issue; the time (days) between the creation of an issues and the registration of their comments was also calculated. These data were calculated to perform predictions of issue success considering periods of time.

\subsection{Feature extraction}

Textual descriptions of comments and issues (which have lexical, syntactic and semantic features) are used in this work to perform prediction tasks. Sequences (n-grams) of words are usually used as lexical features. Syntactic features relate to the structure of texts. Part of Speech (PoS) tags are morphological features that are used to categorize a word in accordance with its syntactic function (nouns, adjectives, verbs, etc.), so they are often used to define syntactic features. Semantic features relate to the meaning of words in a text.
The following features were selected to be extracted from the preprocessed descriptions of issues and comments.

\begin{itemize}
\item	Word n-grams. Words and sequences of words varying from 2 to 10 words were considered.
\item PoS tags n-grams. PoS tags and sequences of PoS tags varying from 2 to 5 were considered.

\end{itemize}

When texts are used to construct models that are used for machine learning classifiers to perform prediction tasks, they must be transformed into numerical vectors that can be understood by the classifiers. Text vectorization is performed based on defined features and weighting schemes. The following weighting schemes were selected to vectorize the descriptions of issues and comments.

\begin{itemize}
\item Term frequency (TF). Score representing the number or occurrences of a term in a document \citep{Jones1972}. 
\item 	Term frequency - Inverse document frequency (TF-IDF). Numerical statistic that is intended to reflect how important a word is to a document in a collection or corpus \citep{Rajaraman2011, Wu2008}. 
\end{itemize}

Experiments were performed using the TF weighting scheme, and then, experiments were also performed using the TF-IDF weighting scheme. This was performed with the aim of comparing which weighting scheme provided the best results.
In both types of experiments, the defined features were considered, so the number of occurrences and the importance of each word n-gram and each PoS tag n-gram were used to perform predictions of issue success.

The feature extraction process was performed using the Python programming language and the scikit-learn library for Python \citep{Manning:1999, Scikitlearn}.
The \textit{CountVectorizer} functionality of the scikit-learn library was used to convert the texts (descriptions of comments and issues) to a matrix of token counts considering the TF weighting scheme and the defined features. The \textit{TfidfVectorizer} functionality was also used to convert the texts to a matrix of TF-IDF features. 

\subsection{Training}
\subsubsection{Algorithm selection}
\label{subsub:algorithmSelection}
The classifiers are machine learning algorithms that are used to predict the classes of given objects. The following algorithms were selected to predict successful and unsuccessful issues.
\begin{itemize}
\item	Multinomial Naïve Bayes (\textbf{MNB}). It is a Naïve Bayes algorithm variant that can be used to classify texts.
\item	Logistic Regression (\textbf{LR}). It is a linear model for classification in which the probabilities describing the possible outcomes of a single trial are modeled using a logistic function.
\item	Support Vector Classifier (\textbf{SVC}). It is a Support Vector Machine (SVM), which is a non-probabilistic binary linear classifier.
\item	Decision Tree Classifier (\textbf{DTC}). It is tree-like model used to perform classifications.
\item	MLP Classifier (\textbf{MLPC}). It is an Artificial Neural Network (ANN) model to perform predictions.
\item	Random Forest Classifier (\textbf{RFC}). It is an ensemble method that uses various decision tree classifiers.
\item	Gradient Boosting Classifier (\textbf{GBC}). It is an ensemble method that supports both binary and multi-class classification.
\end{itemize}

The above algorithms are implemented in the scikit-learn library for Python \citep{Scikitlearn}. The algorithms were selected to identify those with the best results and to compare the results of performing predictions of issues success with results of related works that reported the use those classifiers (the works of \cite{Guo2010} and \cite{SHIPPEY2019} reported the use of LR classifier, the works of \cite{DiSorbo2019} and \cite{Menzies2007} reported the use of Naïve Bayes classifier, and \cite{Hall2011} reported the use of Naïve Bayes and LR classifiers). In addition to the classifiers used in related works, ensemble methods (RFC, which is an averaging method, and GBC, which is a boosting method) were selected to compare results with results of non-ensemble methods such as DTC. Probabilistic classifiers (such as MNB) and non-probabilistic methods (such as SVC) were selected to determine which type of classifiers performed better. A Multi-Layer Classifier (MLPC) was also selected due to its capability to learn non-linear models and its capability to learn models in real-time \citep{Scikitlearn}.

A detailed description of the above algorithms and the parameters that were used to perform experiments are presented in \ref{appClassifiers}.

\subsubsection{Description of experiments}

Experiments were designed to answer the research questions regarding the usefulness of texts (descriptions of issues and comments) to predict issue success. The experiments were performed using the previously selected features, weighting schemes and machine learning algorithms. Experiments were also performed by issue type (bugs, improvements and new features) with the aim of analyzing the differences when predicting the success of each issue type. 
The prediction of issue success over time is studied in this work, so experiments were designed considering the issue resolution time and the time between the creation of issues and the date of reporting of their comments. The resolution time of most of the issues varied from 1 to 3,500 days, so periods of time in this interval were considered. Table \ref{table:variables} shows a summary of the variables that were used to perform the designed experiments. In each experiment, predictions of issue success were performed using an algorithm (A), feature (F), weighting scheme (W), descriptions of issues of a specific type (T) that were resolved in more than a specific number of days (N), and descriptions of comments that were registered in the specified number of days (N) after the creation of their respective issues.

\begin{table}[hbt!]
\small
\begin{center}
 \begin{tabular}{| p{2.2cm} | p{2.2cm}| p{2.8cm} | p{2cm}| p{2.2cm} |}
 \hline
 \textbf{Issue types (T)} & \textbf{Algorithms (A)} & \textbf{Periods of time in days (N)} & \textbf{Features (F)} & \textbf{Weighting schemes (W)} \\ [0.5ex] 
 \hline\hline
 {Bugs, Improvements and New features} & {MNB, LR, SVC, DTC, MLPC, RFC, GBC} & {1, 5, 10, 20, 30, 40, 50, 60, 70, 80, 90, 100, 150, 200, 250, 300, 350, …, 3500. A total of 80 measures of time were considered.} & {Word n-grams, PoS tag n-grams} & {TF, TF-IDF}\\ 
 \hline
\end{tabular}
\caption{Summary of variables for the designed experiments}  
\label{table:variables}
\end{center}
\end{table}

A total of 6,720 experiments were designed and conducted (combining three issue types, seven classifiers, 80 periods of time, two weighting schemes and two feature types) using the selected issues and comments in Table \ref{table:generalData}. For each experiment, the following steps were performed using Python and the scikit-learn library \citep{Scikitlearn}.
\begin{itemize}
\item The description of each issue and the descriptions of its comments were joint into a single text. This was performed because several comments of an issue can provide more information about the progress of the issue than individual comments and, thus, improve the prediction of issue success.
\item Feature extraction was performed to vectorize the selected texts considering the defined features and weighting schemes.
\item The seven algorithms in section \ref{subsub:algorithmSelection} were trained on the vectorized texts to produce prediction models. The texts were divided as follows: 75\% was included in the training set and 25\% in the test set, which was used to evaluate the results. When the number of successful issues was greater than the number of unsuccessful issues (or vice versa), the training set was balanced to perform experiments with the same number of texts in each class using under-sampling balancing (reducing the size of the bigger class). The training set was balanced to represent both classes equally. Experiments with balanced test sets were performed to calculate the accuracy measure (which is sensitive to unbalanced classes) and then, experiments with unbalanced test sets were performed to calculate precision, recall and F1-score in a realistic way (considering the real distribution of issues by periods of time).
\end{itemize}

The above steps were also performed considering the issues of projects in Table \ref{table:projectData}. These experiments were performed considering a balanced training set and an unbalanced test set.

\subsection{Evaluation}

The most used metrics to evaluate the results of machine learning algorithms are based on the number of “True" (correctly predicted) and “False" (incorrectly predicted) values of two generic classes: “Positive” and “Negative”. According the issue resolution tags (Figure \ref{fig:success}), issues (bugs, improvements and new features) are categorized in two general classes (successful and unsuccessful), which can correspond to the “Positive” and “Negative” classes respectively. Based on this, the values that can be used to measure the performance of machine learning algorithms are: True Positives (number of correctly predicted successful issues), False Positives (number of incorrectly predicted successful issues), True Negatives (number of correctly predicted unsuccessful issues), and False Negatives (number of incorrectly predicted unsuccessful issues). These values are summarized in the confusion matrix in \ref{appEvaluation}.
Precision, accuracy recall and F1 score (usually calculated to evaluate the performance of machine learning classifiers) were calculated in each experiment and are described in \ref{appEvaluation}.

\section{Results}
\label{S:5}
In general, PoS tag n-grams were not useful to predict issue success, and the best results were obtained using the TF-IDF weighting scheme and word n-grams as features. The results of predicting issue success considering three issue types, 80 periods of time, seven classifiers, the weighting scheme TF-IDF, and word n-grams are described in detail in the following sections.

\subsection{ Prediction of issue success}

\begin{figure}[ht]
\centering\includegraphics[width=1.0\linewidth]{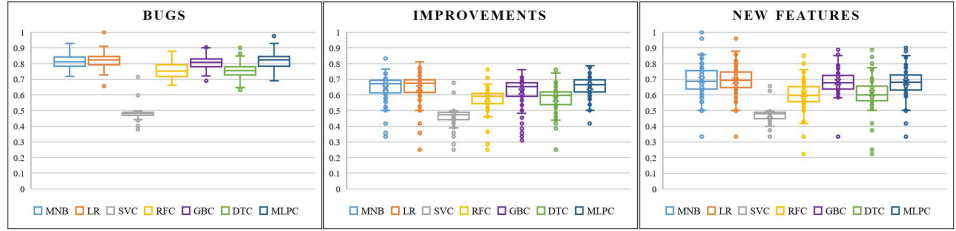}
\caption{Accuracy measurements of predictions of issue success by issue type and classifier}
\label{fig:accuracy}
\end{figure}

Accuracy of predictions of issue success varied for each issue type (bugs, improvements and new features): the accuracy was from 0.38 to 1.0 in experiments with bugs, from 0.25 to 0.83 in experiments with improvements and from 0.22 to 1.0 in experiments with new features. The predictions of bug success were more accurate than the predictions of the success of improvements and new features. The most accurate algorithms were LR, MNB, GBC and MLPC, and the least accurate was SVC. Figure \ref{fig:accuracy} graphics the accuracy measurements of predicting issue success by issue type. Each of the 21 box plots in Figure \ref{fig:accuracy} summarizes the accuracy measurements of 80 experiments (corresponding to 80 periods of time) and show the minimum, maximum, first quartile, third quartile, median and mean of the accuracy measurements that were achieved for a specific machine learning classifier. Descriptive statistics of accuracy measurements are detailed in \ref{appStatistics}.

\begin{figure}[hbt!]
\centering\includegraphics[width=1.0\linewidth]{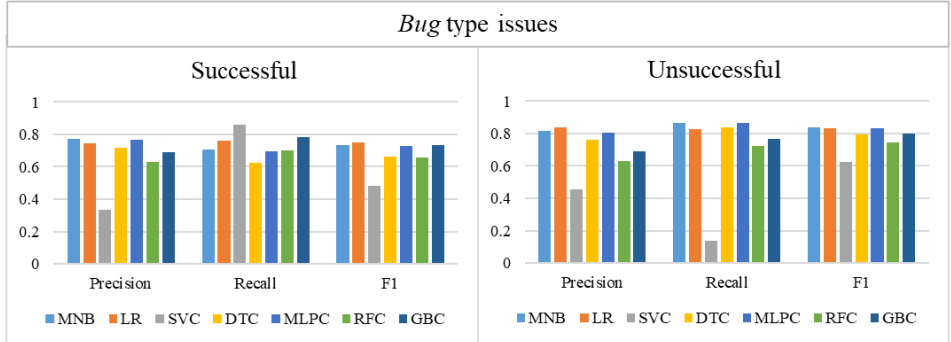}
\caption{Average performance values of predictions of successful and unsuccessful issues (bug type)}
\label{fig:performanceBugs}
\end{figure}

\begin{figure}[hbt!]
\centering\includegraphics[width=1.0\linewidth]{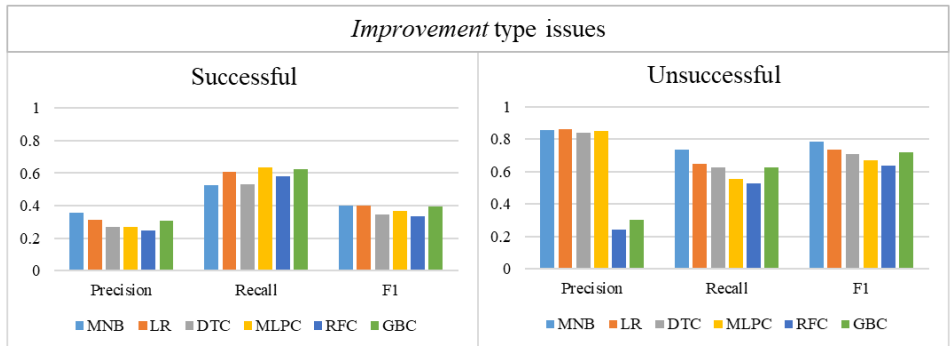}
\caption{Average performance values of predictions of successful and unsuccessful issues (improvement type)}
\label{fig:performanceImpr}
\end{figure}

\begin{figure}[hbt!]
\centering\includegraphics[width=1.0\linewidth]{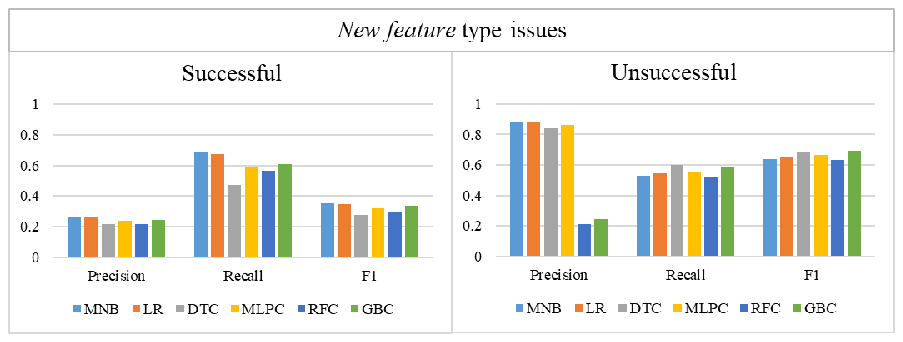}
\caption{Average performance values of predictions of successful and unsuccessful issues (new feature type)}
\label{fig:performanceNewF}
\end{figure}

The precision, recall and F1 score of predicting issue success varied from 0.0 to 1.0 for all the issue types (bugs improvements and new features). The mean precision of predicting unsuccessful bugs and the mean precision of predicting successful bugs were similar.
Figures \ref{fig:performanceBugs}, \ref{fig:performanceImpr} and \ref{fig:performanceNewF} show the mean performance results of predicting issue success by issue type and classifier. The mean performance measurements (precision, recall and F1) of predicting successful and unsuccessful bugs were  higher than the mean performance measurements for predicting improvements and new features. In most of the experiments, SVC achieved the worst results and in most of experiments with improvements and few features, SCV did not predicted any successful or any unsuccessful issue, so its performance measures were not included. Four tables in \ref{appStatistics} present the maximum, minimum, mean, variance and standard deviation of the performance measurements (accuracy, precision, recall and F1 score) by issue type and classifier, and each value was calculated from the results of 80 experiments using the data in Table \ref{table:generalData}.

In general, the best classifiers were MNB, LR, GBC and MLPC (which results were statistically similar) followed by RFC and DTC. The worst classifier was SVC and its results were statistically different to the results of the other classifiers. The performance of MNB and LR classifiers were the best (they performed predictions in less than one second), followed by RFC, DTC and MLPC (they performed predictions in less than one minute), and GBC (with less than two minutes to perform predictions). SVC performed predictions in about nine minutes, so its performance was the worst. The performance (time to perform predictions) of classifiers were measured in an Intel\textsuperscript{\textregistered} CORE i5 7th generation processor in a Windows 10 (64 bits) system.

\subsection{Variation of predictions of issue success over time}

\begin{figure}[hbt!]
\centering\includegraphics[width=1.0\linewidth]{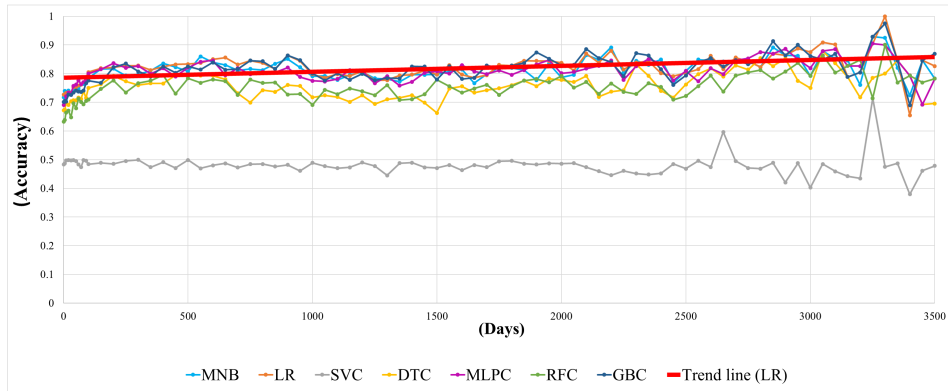}
\caption{Accuracy for predicting bug success over time}
\label{fig:timeBugs}
\end{figure}

\begin{figure}[hbt!]
\centering\includegraphics[width=1.0\linewidth]{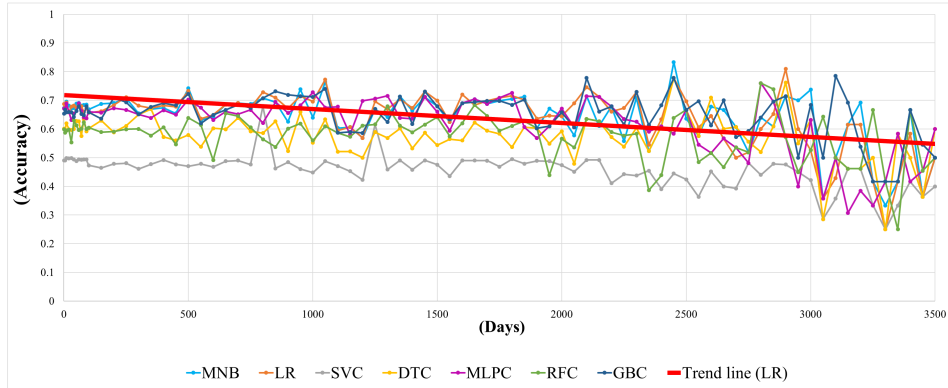}
\caption{Accuracy for predicting improvement success over time}
\label{fig:timeImpr}
\end{figure}

\begin{figure}[hbt!]
\centering\includegraphics[width=1.0\linewidth]{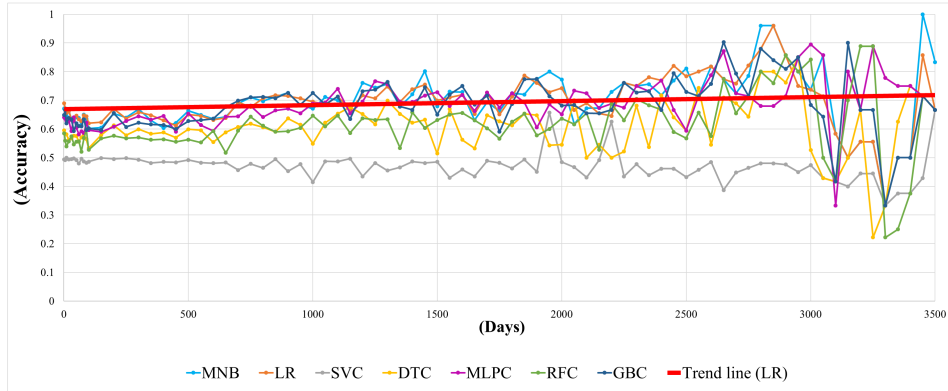}
\caption{Accuracy for predicting new feature success over time}
\label{fig:timeNewF}
\end{figure}

Predictions of issue success were performed considering 80 measures of time in days (Table \ref{table:variables}), which indicate the first \textit{N} days since an issue was created. For each period of time, experiments were performed considering only comments that were created in such period of time and issues that had not been resolved yet. Figures \ref{fig:timeBugs}, \ref{fig:timeImpr} and \ref{fig:timeNewF} show the accuracy for predicting issue success considering the time measures and seven machine learning classifiers.
The accuracy of predicting bug success (Figure \ref{fig:timeBugs}) tended to increase as larger periods of time were considered. This is represented by the trend line in Figure \ref{fig:timeBugs}. The increment on the accuracy of most of the classifiers (except for the SVC) was particularly clear in the first 500 days. The accuracy of predicting the success of improvements (Figure \ref{fig:timeImpr}) seemed to increase until 500 days, after greater periods of time, the accuracy tended to decrease. This decrement is represented by the trend line in Figure \ref{fig:timeImpr}.
The accuracy of predicting the success of new features seemed to be little variant in the first measures of time, and after 1000 days, it seemed to be irregular; however, the accuracy of predicting the success of new features tended to increase as the number of days that were considered increased (Figure \ref{fig:timeNewF}).
In general, the accuracy of classifiers seemed to be variant when considering periods of time of more than 1500 days.

\begin{figure}[hbt!]
\centering\includegraphics[width=1.0\linewidth]{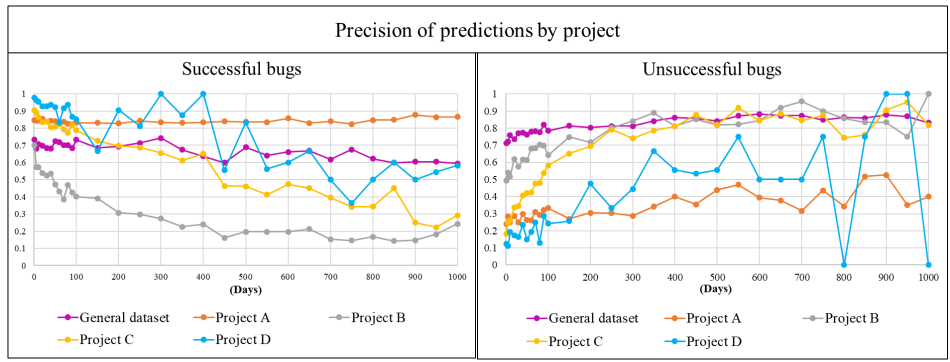}
\caption{Precision for predicting bug success over time by project using LR classifier}
\label{fig:precisionProjects}
\end{figure}

\begin{figure}[hbt!]
\centering\includegraphics[width=1.0\linewidth]{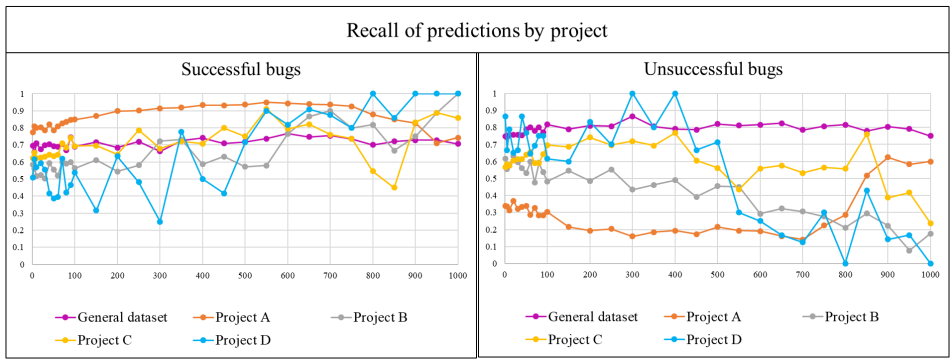}
\caption{Recall for predicting bug success over time by project using LR classifier}
\label{fig:recallProjects}
\end{figure}

Precision, recall and f1 score were calculated in experiments by periods of time using data of individual projects in Table \ref{table:projectData}. These results were compared with the results using the general dataset in Table \ref{table:generalData}.
Results showed that precision and recall of the general dataset was better than most of the individual projects as shown in Figure \ref{fig:precisionProjects} and Figure \ref{fig:recallProjects}. Precision of predicting successful bugs, and recall of predicting unsuccessful bugs using issues and comments of Project C were very similar to the results using the general dataset; however, recall and precision could vary depending on the distribution and number of issues and comments.

Precision and recall of predicting bug success using data from Project A were less variant than results using data from the other projects. This is explained because project A have more issues and comments than the other projects, and predictions were less variant.
Most of predictions of successful bugs using data from specific projects achieved more than 70\% of precision in the first measures of time. Precision of predicting unsuccessful bugs, and recall of predicting successful and unsuccessful bugs tended to increase as larger periods of time were considered using both, the general dataset and data from the specific projects.

\begin{figure}[hbt!]
\centering\includegraphics[width=0.7\linewidth]{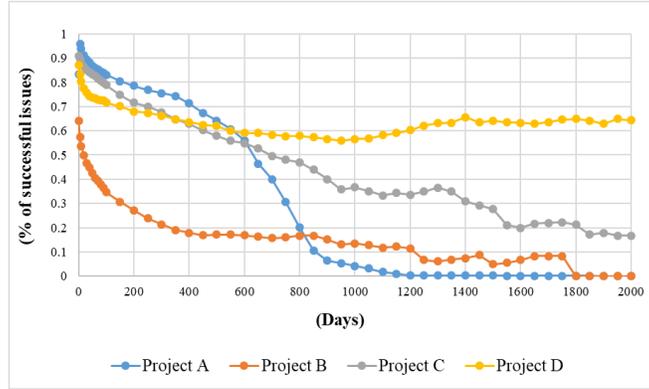}
\caption{Percentage of successful issues over time by project}
\label{fig:successTime}
\end{figure}

As shown in Figure \ref{fig:successTime}, the percentage of successful issues decreased as higher periods of time were considered. Correlations between the percentage of successful issues and time were calculated by project using the Pearson’s correlation coefficient \citep{pearson1895note, Stigler}. The correlation coefficient was -.936 for project A, -.88 for project B, -.98 for project C and -.56 for project D. These correlations indicate that the development of software tasks was more effective in the first periods of time and that more unsuccessful issues existed at the end of the projects.

\subsection{Relevant information for predicting issue success}

The most relevant features (characteristics of descriptions of issues and comments) for predicting issue success were identified in each of the 80 experiments by issue type using the general dataset. The features were ranked according to their importance in each experiment according to their weights (numeric values that were obtained in the vectorization process). Tables \ref{table:featSuccess} and \ref{table:featUnsucc} show the most relevant features and indicate the times the features were among the 100 most relevant features and the average ranking of each feature. Only the relevant features for predicting issue success that were identified in experiments using the general dataset and that were also identified in experiments using data of the specific projects were included in Tables \ref{table:featSuccess} and \ref{table:featUnsucc}. In general, the most relevant features for predicting issue success were unigrams of words.

\begin{table}[hbt!]
\footnotesize
  \begin{tabular}{ |c||c|c|c|c|c|c|c|c|c|  }
  \hline
  {}&
  \multicolumn{3}{|c|}{\textbf{BUGS}}&
  \multicolumn{3}{|c|}{\textbf{IMPROVEMENTS}}&
  \multicolumn{3}{|c|}{\textbf{NEW FEATURES}}\\
   
  \hline
  {\textit{\#}} & {\textit{Feature}}&{\textit{N}}&{\textit{AR}}& {\textit{Feature}}&{\textit{N}}&{\textit{AR}}& 
   {\textit{Feature}}&{\textit{N}}&{\textit{AR}}\\
   \hline
1&patch&80&26.97&currently&78&34.92&used&79&26.81\\
2&file&80&36.36&property&75&34.42&message&76&27.77\\
3&problem&80&55.61&added&74&43.46&make&76&52.42\\
4&fixed&79&30.31&created&74&64.92&set&75&55.14\\
5&bug&79&47.73&methods&71&47.74&security&75&37.35\\
6&document\_type&78&48.27&class&69&51.54&attribute&72&63.73\\
7&attachment&70&54.25&exception&69&54.51&component&66&57.52\\
8&code&70&82.33&build&69&66.87&implement&65&89.02\\
9&thanks&67&72.46&allow&66&69.96&web&65&98.81\\
10&line&65&83.52&object&64&61.91&based&60&136.97\\

   \hline

  \end{tabular}
 
 \begin{tablenotes}
      \small
      \item N: Number of experiments in which the feature was among the 100 most relevant features. AR: Average ranking of the feature considering 80 experiments.
    \end{tablenotes}
\caption{Most relevant features for predicting successful issues by issue type}  
\label{table:featSuccess}
\end{table}

\begin{table}[hbt!]
\footnotesize
  \begin{tabular}{ |c||c|c|c|c|c|c|c|c|c|  }
  \hline
  {}&
  \multicolumn{3}{|c|}{\textbf{BUGS}}&
  \multicolumn{3}{|c|}{\textbf{IMPROVEMENTS}}&
  \multicolumn{3}{|c|}{\textbf{NEW FEATURES}}\\
   
  \hline
   {\textit{\#}} & {\textit{Feature}}&{\textit{N}}&{\textit{AR}}& {\textit{Feature}}&{\textit{N}}&{\textit{AR}}& 
   {\textit{Feature}}&{\textit{N}}&{\textit{AR}}\\
   \hline
1&url\_spec.&80&3.27&url\_spec.&80&2.08&code\_spec.&80&4.32\\
2&review&80&26.71&version\_spec.&80&6.96&like&80&8.68\\
3&project&80&30.91&code&80&12.43&need&80&14.32\\
4&error&80&38.21&make&80&18.91&project&80&17.52\\
5&issue&78&44.62&need&80&20.23&using&80&22.61\\
6&test&75&62.72&using&80&20.25&patch&80&25.66\\
7&user&70&91.62&issue&80&20.28&feature&80&29.08\\
8&click&69&64.92&new&80&20.60&just&80&29.43\\
9&spring&66&66.08&project&80&23.58&version&80&30.07\\
10&would&60&97.62&add&80&23.67&nice&79&38.42\\

   \hline
  \end{tabular}
 
 \begin{tablenotes}
      \small
      \item N: Number of experiments in which the feature was among the  100 most relevant features. AR: Average ranking of the feature considering 80 experiments.
    \end{tablenotes}
\caption{Most relevant features for predicting unsuccessful issues by issue type }  
\label{table:featUnsucc}
\end{table}

\begin{figure}[hbt!]
\centering\includegraphics[width=1.0\linewidth]{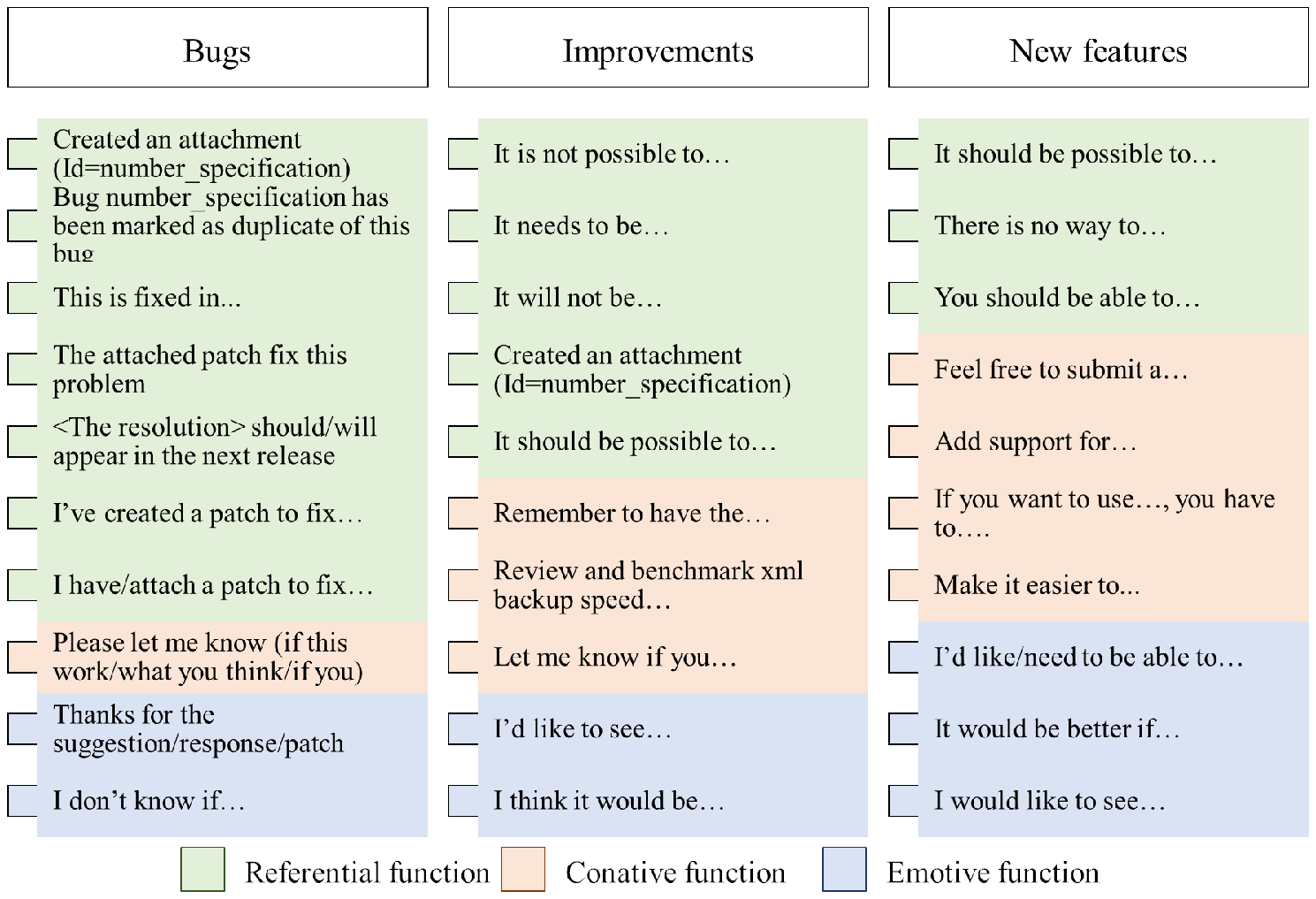}
\caption{Most relevant phrases for predicting successful issues}
\label{fig:phrasesS}
\end{figure}

\begin{figure}[hbt!]
\centering\includegraphics[width=1.0\linewidth]{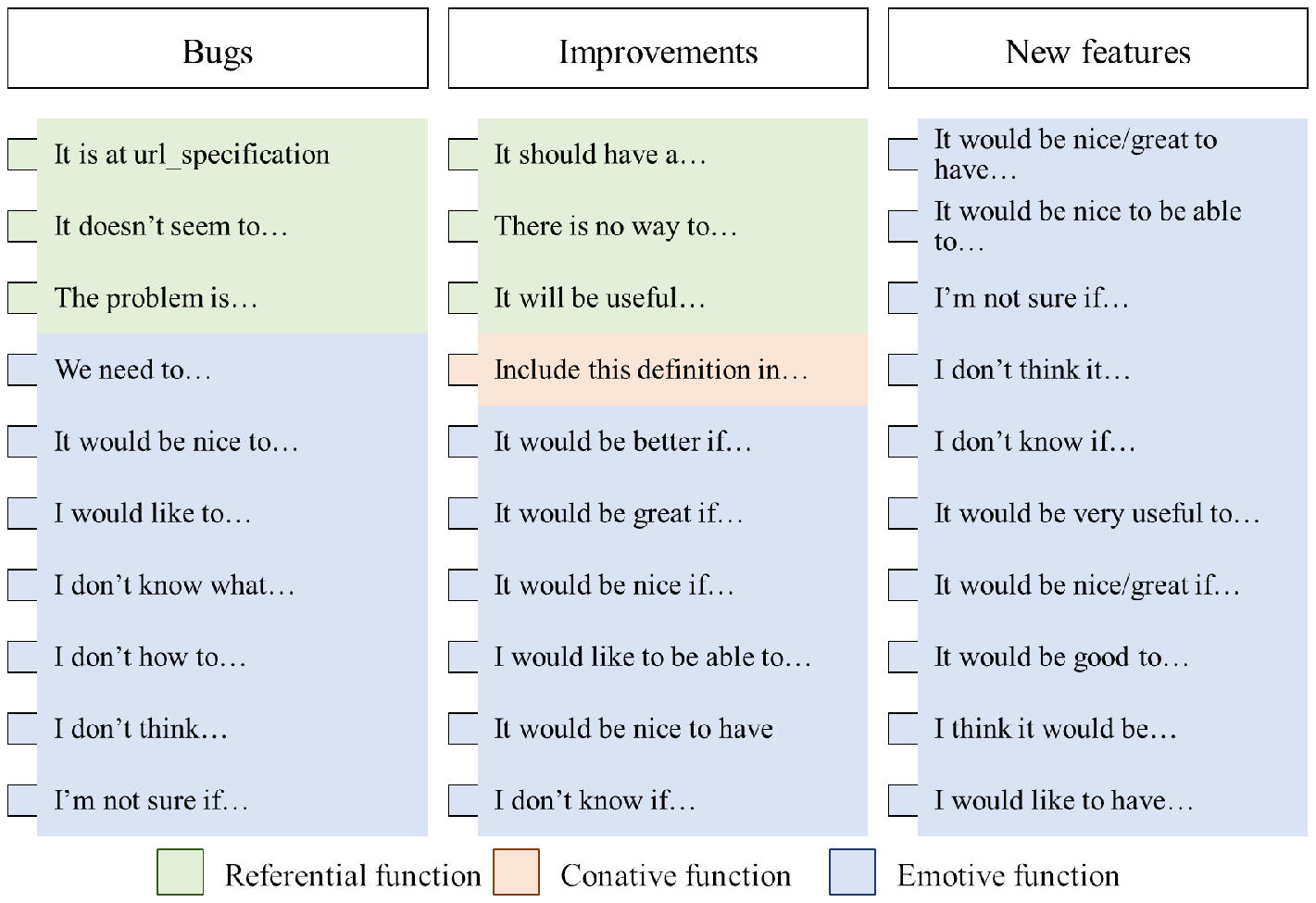}
\caption{Most relevant phrases for predicting unsuccessful issues}
\label{fig:phrasesU}
\end{figure}

Most of the relevant features in Tables \ref{table:featSuccess} and \ref{table:featUnsucc} are nouns, and most of them are technical concepts that relate to the context of the study (software development).
The most relevant phrases for predicting issue success were also identified. Figures \ref{fig:phrasesS} and \ref{fig:phrasesU} show 10 of the most representative and relevant phrases for predicting issue success by issue type. The phrases were classified according to their function (or purpose) with the aim of understanding  the communication functions associated with issue success in software development. The phrases in Figures \ref{fig:phrasesS} and \ref{fig:phrasesU} were classified according to the following categories of communication functions proposed by \cite{jakobson1963}: the referential function indicates that a phrase describes a situation, context, or state; the conative function is used in imperative sentences and indicates that a phrase is intended to change people behavior; and the emotive function indicates that a sentence expresses emotions, thoughts, wishes, etc.
Most of the relevant phrases for predicting successful issues were referential phrases, particularly for predicting successful bugs and improvements (Figure \ref{fig:phrasesS}). The purpose of many of these phrases was to provide technical information. Some conative phrases and few emotive phrases were identified as useful for predicting successful issues.
Most of the relevant phrases for predicting unsuccessful issues were emotive phrases, which are not intended to provide technical information. Few referential phrases and only one conative phrase, were identified as relevant phrases for predicting unsuccessful issues (Figure \ref{fig:phrasesU}).

PoS tags n-grams were not identified as relevant features for the prediction of issue success; however, some of the most representative PoS tags n-grams were identified and are the following: “noun + noun”, “determiner + noun”, “noun + preposition”, and “noun + verb”. 

\section{Discussion}
\label{S:6}

\subsection{Answers to research questions}

\subsubsection{Are textual descriptions of issues and comments from Jira ITSs useful to predict issue success in software projects?}

In general, descriptions of issues and comments were useful for predicting the success and failure of issues, and particularly, descriptions of bugs and their comments were very useful. 
The best classifiers achieved an accuracy and precision between 80\% and 100\% for predicting issue success in many experiments. Since day one, some classifiers achieved an accuracy and precision of more than 70\% for predicting successful and unsuccessful bugs, which indicates that the success or failure of more than 70\% of bugs can be predicted correctly the day a bug is reported using its description and comments. Since day 30, some classifiers achieved more than 80\% of precision and recall when predicted bug success.

The evaluation of prediction results depends on the type of the prediction problem. In this work, the prediction of the success or failure of issues is aimed to make decisions early and take actions to ensure the successful completion of software tasks. If it is predicted that an issue will be unsuccessfully performed, actions to ensure its success can be taken, and if the prediction is wrong, extra actions to ensure the successfully completion of the issue do not affect software quality. Based on the above, more than 80\% of precision and recall of predicting issue success can be considered as good because at least 80\% of the issues that will be unsuccessfully performed can be identified after 30 days and can be managed to avoid their failure, and at least 70\% of the issues that will be unsuccessfully resolved can be predicted since the day of their report. The time for resolving an issue is 500 days on average, so the success or failure of most of the issues can be predicted with more than 80\% of precision in the 10\% of the time that takes their resolution. This indicates that, since the early activities involved in an issue resolution, early decisions can be taken and actions can be performed to ensure the success of the issue.

\subsubsection{What kind of information is useful to predict issue success?}

The most relevant features for the prediction of issue success were single words related to software development processes. The words “added", “created", “fixed” and “used” were relevant words for predicting successful bugs, improvements and new features. These words were used in phrases to report an action intended to resolve issues.

The relevant words for predicting unsuccessful issues included “url\_specification” and “code\_specification”, which refers to URLs and code fragments that were used to indicate that specific software modules or features needed to be fixed. The words “project”, “error” and “review" were also relevant for predicting unsuccessful issues. These words were used to report specific software concerns related to issues that were unsuccessfully addressed. 

The phrases that were more relevant for predicting issue success were referential phrases. Phrases that provided technical information, described situations, and reported work progress, were strongly related to successful bugs. This indicates that during the correction of bugs that will be successfully resolved, people share technical information that is useful for the correction of bugs (such as attachments, patches, resolutions and code) and report advances on the correction of bugs, which highlights objective communication. 

In general, most of the relevant phrases for predicting successful issues were referential and conative phrases. This suggest that when issues are being addressed successfully, people report work, describe context, suggest work and exhort people to perform work through comments. In contrast, the most relevant phrases for predicting unsuccessful issues were emotive phrases, which express feelings, emotions, suggestions and thoughts. This indicates that when issues that will be unsuccessfully attended are being developed, people express personal aspects and concerns about work that is not being developed rather than technical information. The phrases “I don’t know…” and “I’m not sure…”, which were identified as relevant phrases for predicting unsuccessful issues, indicate misunderstanding and unclarity. This highlights that an ineffective communication may relate to unsuccessful issues because effective communication implies clarity and that the intended meanings equals the perceived meanings \citep{Schermerhorn2002}.

Some referential phrases such as “The problem is…” or “There is no way to…”, were also identified as relevant phrases for predicting unsuccessful issues, but they indicated that something was going wrong, and that few advances on issue resolution existed.  

\subsubsection{How does the prediction of issue success vary over time?}

Predictions of issue success were performed considering 80 periods of time. In general, accuracy, and some  precision and recall measures of issue success predictions tended to increase as larger periods of time were considered. 
Predictions in first periods of time are most relevant because people can know early whether issues will be  successfully resolved or not. 
Accuracy for predicting improvement success tended to increase until 550 days, and after that, it tended to be irregular and decrease in the last periods of time. This indicates that, in general, the information related to bugs and new features is very useful for predictions in most of the periods of time, and that the first comments that were registered during the development of improvements are the most useful for predicting issue success. This highlights that the information communicated during the resolution of bugs and new features is related to the success or failure of issues during all the resolution process, and that only the initial information that is communicated during the development of improvements is related to the resolution result. Improvements are perfective maintenance tasks, and about 86\% of their comments are written until day 500, and after that, new useful information (such as referential phrases) rarely arise. This indicates that after day 500, few information is provided because few issues remain unresolved. This causes that accuracy and precision of predictions decrease after day 500.

Accuracy of experiments for predicting three types of issues (bugs, new features and improvements) was less variant considering the first measures of time than the accuracy of experiments considering the last measures. This can be explained because the number of texts decreased in the last measures of time (in the experiments, issues that have not been resolved yet were considered, so in experiments considering the initial measures of time, predictions were performed with more texts than in the last predictions when periods of time of more than 2000 days where considered), and few texts could make the results of predictions unstable and variable.

Bugs, which are activities of corrective maintenance, were attended in an approximate time of 480 days, improvements in 460 days and new features in 550 days. This indicates that an accurate prediction of bug success and improvement success can be performed earlier than an accurate prediction of the success of new features because the progress on the resolution of bugs and improvements can be reflected in comments earlier than in comments of new features.

\subsubsection {How does the prediction of issue success vary with respect to the three issue types (bugs, improvements and new features)?}

In general, predictions of bug success were more accurate and precise than the predictions of the success of improvements and new features. Bugs are relevant software tasks because affect software outcomes, so when a defect or failure is identified, people must begin working on fixing it. Bugs are fixed in less time than new features, so an important number of comments of bugs that are reported in the first periods of times are useful for predicting bug success. This may indicate that the communication is more efficient when bugs are attended because their comments include important referential information for resolving them.

Improvements and new features are perfective and adaptive maintenance tasks respectively, which are not often a priority in software development, so few people are sometimes assigned to attended them. Results shows that messages tend to be conative and poorly referential when improvements and new features are developed. This may explain that the messages that are communicated during the development of new features and improvements are less useful to perform predictions of issue success than messages communicated during bug resolution.

\subsection {Related work}
In the present work, some of the classifiers that achieved the best results were Multinomial Naïve Bayes (which is a variation of Naïve Bayes) and Logistic Regression (LR). Other works that focus on predicting other software development outcomes (such as fault prediction), have found that these two algorithms performed well; however, almost no works on predicting issue success have been conducted. \cite{Menzies2007} showed that Naïve Bayes provided better performance for fault prediction than other algorithms. The results of \cite{Hall2011} showed that he models that performed well tend to be based on simple modeling techniques such as Naive Bayes or Logistic Regression. \cite{SHIPPEY2019}  identified code features for software defect prediction and found that Logistic Regression achieved good results.

Some works highlighted the importance of comments for predictions in software projects. In the work of \cite{Choetkiertikul2018}, the number of comments was one of the most relevant features. \cite{VALDIVIAGARCIA2018} concluded that the description and comments of bugs were the most important factors to predict bugs that block the fixing of other bugs. The results of \cite{DiSorbo2019} demonstrated that it is possible to predict whether an issue will be closed as a “won't fix” using textual features from titles and descriptions of issues; and that Naïve Bayes had a precision of 0.795. In the present work, the mean precision for predicting bug success was over 0.80. The work of \cite{DiSorbo2019} is similar to the present work; however, the present work studies three issue types and many types of issue resolutions classified in successful and unsuccessful. In the present work, descriptions of comments were also used to predict issue success (not only titles and descriptions of issues), and relevant phrases for predicting issue success (associated to communication functions) were identified. The present work studied issues from Jira ITSs and one of its most important contribution is the performing of predictions considering periods of time to study how predictions vary according to how long an issue has been opened. The present work provides evidence that prediction of issue success can be performed using data of open (or not resolved) issues. In addition, the present work also contributes on studying the usefulness of seven machine learning classifiers to perform issue success predictions.

In the work of \cite{Guo2010}, 68\% of precision and 64\% of recall were achieved when predicting Windows 7 bug fixes; in the present work (using only descriptions of issues and comments) the mean precision and recall of the best classifiers were greater than 80\%.  \cite{Guo2010} recommended to train employees to write high-quality bug reports and improve communication and trust amongst people. The present work provides evidence on the importance of communicated comments for issue success.

The results of \cite{Murguia} showed that perfective maintenance (including the development of improvements) is on average faster than corrective maintenance (including the resolution of bugs); and corrective maintenance is on average faster than adaptive maintenance tasks (including the attention to new features). The results of the present work confirm the results of \cite{Murguia} because it was found that improvements are resolved in an average time of 460, bugs in 480 days and new features in 550 days.

\subsection {Implications for practice}

This study benefits the software development community because describes a simple way to predict the success of software tasks based on text (descriptions of issues and comments) analysis. Particularly, organizations that use ITSs to manage their projects can perform early predictions of issue success, reduce time in the development processes, increase software quality and customer satisfaction, and improve the organizational productivity. Early predictions of issue success can help project managers to distribute resources and attend critical issues that are likely to fail; developers can reduce development time because they could know early whether issues will be attended and take actions without having to wait until the resolution of a tasks to perform other software development activities; and customers can know early if their requirements will be satisfied.

This work provides evidence on how to use texts (description of issues and comments) to predict issue success and evidence about the best machine learning classifiers to  perform this task. This work highlights the most useful features and kind of texts to perform predictions of issue success with good accuracy and precision considering periods of time with the aim of helping organizations to perform accurate and precise predictions.

This work provides evidence that aspects of organizational performance (such as issue success) can be predicted using textual information of interactions among people (not always, the use of technical information is required), so organizations can perform predictions in a simple way. Organizations dedicated to software development must consider a frequent analysis of data from ITSs as an important strategy for improving software development tasks. 

Based on the results, authors of the present work recommend the following: include referential information on descriptions of issues and comments to increase the quality of texts and the probability of issue success; encourage objective, frequent and open communication among people involved in a software project; perform predictions of issue success as soon as possible (enough information can be obtained to perform highly accurate and precise predictions after 30 days an issue was registered); implement an automatic process to predict issue success that can be updated continuously with recent data to perform predictions; use MNB and LR when texts are considered to predict issue success due to their precise and accurate results and the time they require to perform predictions.

\subsection{Threats to validity}

In this section, the threats that have an impact on the validity of the results are discussed, including threats to construct validity, threats to internal validity, threats to external validity, threats to conclusion validity, and threats to reliability.

Construct validity refers to the degree to which the operationalization of the measurements in a study actually represents the constructs in the real world \citep{Jedlitschka2008}. In this work, the studied data (issues and comments of real software projects) were directly extracted from four Jira ITSs and were not manipulated before being processed, so the data represent information of the real world. One of the advantages of using data from electronic databases such as ITSs, is that the extracted data is stable and is not influenced by the presence of researchers \citep{Jedlitschka2008}.

Internal validity refers to the extent to which the treatment or independent variable(s) were actually responsible for the effects seen to the dependent variable \citep{Jedlitschka2008}. Experiments were performed using balanced traning sets, including the same number of successful and unsuccessful issues to avoid the effects of a variable training set. In some experiments considering measures of time of more than 2000 days, few data were used to perform experiments, so, in these cases, results may have been influenced by the amount of training data; however, more than 6000 experiments were performed to observe the variation of issue success predictions over time. In addition, data preprocessing was a validated process, so threats related to the effect of manipulations on results were mitigated.

External validity refers to the degree to which the findings of the study can be generalized to other participant populations or settings \citep{Jedlitschka2008}.  In this work, data from four public Jira ITSs that store a considerable number of software projects were selected as data source. The selected Jira ITSs are some of the most used ITSs due to the kind of projects that store, which include the development of software products that are used by a lot of people that are dedicated to software engineering. These software products include software development tools, plugins, libraries, frameworks, programming languages, IDEs, and collaboration tools, and many of them are Apache, Spring and Atlassian products; thus, this study must be of the interest of a lot of people dedicated to software development. The data that were extracted from the ITSs include issues and comments from 588 software projects (representing more than 50\% of the total projects of the selected ITSs), so the studied data are a representative part of the projects that are recorded in public Jira ITSs. In addition, more than 6000 experiments were performed to study the prediction of issue success considering different variables (algorithms, features, weighting schemes, issue types, periods of time) with the aim of considering most of the possible cases and scenarios.

Conclusion validity refers to whether the conclusions reached in a study are correct \citep{Jedlitschka2008}. In this study, conclusions of results were stated considering the limitations and context of this work.

Reliability is concerned with the extent to which the data and the analysis are dependent on specific researchers \citep{Runeson2012}. The studied data were extracted directly from four Jira ITSs, and they were not modified by any researcher. The data analysis was automatic, so did not depend on specific researchers. Most of the experiments were executed several times to provide reliability to the study. The data extraction, data preprocessing, and the experiment execution activities were validated.

\section{Conclusions and future work}
\label{S:7}
In this work, the usefulness of descriptions of issues and comments for predicting issue success was studied. More than 6000 experiments were performed considering seven machine learning classifiers, 80 measures of time and three types of issues (bugs, improvements and new features).

Results showed that descriptions of issues and comments can be used for predicting issue success with good levels of accuracy and precision. Some words that relate to software development were particularly useful for predicting issue success. 

The prediction of issue success can reduce costs in software development and improve software quality because people can identify the issues that will not be successfully addressed. This can be useful for performing actions to avoid issue failure and reduce development time.

More research is needed on predicting issue success, and data form other repositories and tools must be considered. Other aspects of people interaction and human factors must be considered as features for performing predictions in software development.



\newpage
\appendix
\sloppy
\section{Machine learning classifiers}
\label{appClassifiers}
\textbf{Multinomial Naïve Bayes (MNB)}

Naïve Bayes is a probabilistic classifier that uses the Bayes’ theorem, which expresses the probability of an aleatory event knowing conditions that might be related to the event. Naive Bayes is widely used in machine learning due to its efficiency and its ability to combine evidence from a large number of features \citep{Basu2003, Mitchell1997}. Naive Bayes assumes that the value of a particular feature is independent of the value of any other feature, given the class variable \citep{Manning:1999}. Naïve Bayes classifier can be extremely fast compared to more sophisticated methods and they require a small amount of training data to estimate the necessary parameters \citep{Scikitlearn, Zhang2004}. The scikit-learn library implements the Multinomial Naïve Bayes (MNB) classifier, which is a Naive Bayes algorithm variant that is recommended for text classification. The parameters of MNB that were used in this work are the following.

\textit{MultinomialNB (alpha=1.0, class\_prior=None, fit\_prior=True)}

\hfill \break
\textbf{Logistic Regression (LR)}

Logistic Regression (LR) is also known as logit regression, maximum-entropy classification or the log-linear classifier. LR, despite its name, is a linear model for classification rather than regression in which the probabilities describing the possible outcomes of a single trial are modeled using a logistic function \citep{Scikitlearn}. The scikit-learn library implements the LR classifier. The parameters of LR that were used in this work are the following.

\textit{LogisticRegression (C=1.0,  class\_weight=None, dual=False, fit\_intercept=True, intercept\_scaling=1, max\_iter=100, multi\_class='warn',  n\_jobs=None, penalty='l2',  random\_state=None, solver='warn', tol=0.0001,  verbose=0, warm\_start=False)}

\hfill \break
\textbf{Support Vector Classifier (SVC)}

A Support Vector Machine (SVM) is a non-probabilistic binary linear classifier \citep{Garreta2013}. It uses a subset of training points in the decision function (called support vectors), so it is also memory efficient \citep{Seal1967}. SVM is effective in high dimensional spaces; still effective in cases where the number of dimensions is greater than the number of samples \citep{Scikitlearn}. The scikit-learn library implements the Support Vector Classifier (SVC), which is an SVM classifier. The parameters of SVC that were used in this work are the following.

\textit{SVC (C=1.0, cache\_size=200, class\_weight=None, coef0=0.0, decision\_function\_shape='ovr', degree=3, gamma='auto\_deprecated', kernel='rbf', max\_iter=-1, probability=True, random\_state=None, shrinking=True, tol=0.001, verbose=False)}

\hfill \break
\textbf{Decision Tree Classifier (DTC)}

Decision Tree is a tree-like structure that is used for classification and regression. Each internal node of a decision tree denotes a test on an attribute, each branch represents an outcome of the test and each leaf node represents a class label. The scikit-learn library implements the Decision Tree Classifier (DTC), which is capable of performing multi-class classification on a dataset \citep{Scikitlearn}. The parameters of DTC that were used in this work are the following.

\textit{DecisionTreeClassifier (class\_weight=None, criterion='gini', max\_depth=None, max\_features=None, max\_leaf\_nodes=None, min\_impurity\_decrease=0.0, min\_impurity\_split=None, min\_samples\_leaf=1, min\_samples\_split=2, min\_weight\_fraction\_leaf=0.0, presort=False, random\_state=None, splitter='best') }

\hfill \break
\textbf{Multi-Layer Perceptron Classifier (MLPC)}

Artificial Neural Networks (ANN) are computing systems that consist of a set of unities called neurons that are connected to perform classifications or predictions. The MLP Classifier (MLPC) is an ANN model that implements a multi-layer perceptron (MLP) algorithm that trains using Backpropagation \citep{Scikitlearn}. MLPC is implemented in the scikit-learn library and the parameters of MLPC that were used in this work are the following.

\textit{MLPClassifier (activation='relu', alpha=1e-05, batch\_size='auto', beta\_1=0.9, beta\_2=0.999, early\_stopping=False, epsilon=1e-08, hidden\_layer\_sizes=(5, 2), learning\_rate='constant', learning\_rate\_init=0.001, max\_iter=200, momentum=0.9, n\_iter\_no\_change=10, nesterovs\_momentum=True, power\_t=0.5, random\_state=1, shuffle=True, solver='lbfgs', tol=0.0001, validation\_fraction=0.1, verbose=False, warm\_start=False)}

\hfill \break
\textbf{Ensemble Methods}

Ensemble methods combine the predictions of several base estimators built with a given learning algorithm in order to improve generalizability/robustness over a single estimator \citep{Scikitlearn}. Two families of ensemble methods are usually distinguished.

\begin{itemize}
\item \textbf{Averaging methods}, in which the driving principle is to build several estimators independently and then to average their predictions. On average, the combined estimator is usually better than any of the single base estimator because its variance is reduced \citep{Scikitlearn}. The scikit-learn library includes an averaging algorithm based on randomized decision trees: Random Forest Classifier (\textbf{RFC}) \citep{Scikitlearn}. RFC erturb-and-combine techniques \citep{breiman1998ac} specifically designed for trees. This means a diverse set of classifiers is created by introducing randomness in the classifier construction. The prediction of the ensemble is given as the averaged prediction of the individual classifiers \citep{Scikitlearn}. Random Forest is a meta estimator that fits a number of decision tree classifiers on various sub-samples of the dataset and uses averaging to improve the predictive accuracy and control over-fitting \citep{Breiman2001, Seal1967}. The parameters of RFC that were used in this work are the following.

\textit{RandomForestClassifier (bootstrap=True, class\_weight=None, criterion='gini', max\_depth=None, max\_features='auto', max\_leaf\_nodes=None, min\_impurity\_decrease=0.0, min\_impurity\_split=None, min\_samples\_leaf=1, min\_samples\_split=2, min\_weight\_fraction\_leaf=0.0, n\_estimators='warn', n\_jobs=None, oob\_score=False, random\_state=None, verbose=0, warm\_start=False)}

\item	\textbf{Boosting methods}, in which base estimators are built sequentially and one tries to reduce the bias of the combined estimator. The motivation is to combine several weak models to produce a powerful ensemble \citep{Scikitlearn}. Gradient boosting (GB) is a machine learning technique to perform classification and regression tasks using prediction models, typically decision trees. The classifier builds an additive model in a forward stage-wise fashion; it allows for the optimization of arbitrary differentiable loss functions \citep{Seal1967}. Gradient Boosting Classifier (\textbf{GBC}) supports both binary and multi-class classification and is part of the scikit-learn library \citep{Scikitlearn}. The parameters of GBC that were used in this work are the following.

\textit{GradientBoostingClassifier (criterion='friedman\_mse', init=None, learning\_rate=0.1, loss='deviance', max\_depth=3, max\_features=None, max\_leaf\_nodes=None, min\_impurity\_decrease=0.0, min\_impurity\_split=None, min\_samples\_leaf=1, min\_samples\_split=2, min\_weight\_fraction\_leaf=0.0, n\_estimators=100, n\_iter\_no\_change=None, presort='auto', random\_state=None, subsample=1.0, tol=0.0001, validation\_fraction=0.1, verbose=0, warm\_start=False)}
\end{itemize}

\newpage
\sloppy
\section{Measurements for evaluating machine learning algorithms}
\label{appEvaluation}

\begin{table}[hbt!]
\small
  \begin{tabular}{ p{1cm}|c|c|c|  }

  \multicolumn{2}{c}{}&
  \multicolumn{2}{c}{\textit{Predicted values}}\\
  \cline{3-4}
  
   \multicolumn{2}{c|}{}
   & {\textbf{Unsuccessful issues}}& {\textbf{Unsuccessful issues}}\\
  \cline{2-4}
  
  \multirow{2}{4em}{\textit{Actual values}} &
  {\textbf{Successful issues}} & {True Positives (TP)}& {False Negatives (FN)}\\  \cline{2-4}&
  {\textbf{Unuccessful issues}}  &
   {False Positives (FP)}& {True Negatives (TN)}\\

  \cline{2-4}
  
  \end{tabular}
  
\caption{Confusion matrix}  
\label{table:confusionMatrix}
\end{table}

\textbf{Precision}. The ratio of correctly predicted observations in a class to the total predicted observations in such class; \ref{eq:PS} and \ref{eq:PUS} were used to calculate the precision for predicting successful and unsuccessful issues respectively.

\begin{equation}
\label{eq:PS}
Precision_1 = \frac{TP}{TP+FP}
\end{equation}

\begin{equation}
\label{eq:PUS}
Precision_2 = \frac{TN}{TN+FN}
\end{equation}

\textbf{Accuracy}. The ratio of correctly predicted observations to the total observations. 	
\begin{equation}
\label{eq:Ac}
Accuracy=\frac{TP+TN}{TP+FP+FN+TN}
\end{equation}

\textbf{Recall}. The ratio of correctly predicted observations in a class to the all observations in such actual class;  \ref{eq:RS} and \ref{eq:RUS} were used to calculate the recall for predicting successful and unsuccessful issues respectively. 
\begin{equation}
\label{eq:RS}
Recall_1 = \frac{TP}{TP+FN}
\end{equation}
\begin{equation}
\label{eq:RUS}
Recall_2 = \frac{TN}{TN+FP}
\end{equation}
\textbf{F1 score}. The weighted average of precision and recall. 
\begin{equation}
\label{eq:F1}
F1\_score = \frac{2*(Recall * Precision)}{Recall+Precision}
\end{equation}

\newpage
\section{Descriptive statistics of results}
\label{appStatistics}

\textbf{Accuracy results of predicting successful and unsuccessful issues}
\begin{table}[hbt!]
\tiny
\begin{center}
 \begin{tabular}{|c|c|c|c|c|c|c|c|}
 \hline
{} & {MNB} & {LR} & {SVC} & {DTC} & {MLPC} & {RFC} & {GBC} \\ 
 \cline{1-8}
 \multicolumn{8}{|c|}{BUGS} \\ 
 \hline
 Min.&0.72&0.66&0.38&0.66&0.69&0.63&0.69\\
Max.&0.93&1&0.71&0.88&0.9&0.9&0.98\\
Mean&0.81&0.82&0.48&0.76&0.81&0.75&0.81\\
Variance&0&0&0&0&0&0&0\\
Standard deviation&0.04&0.05&0.04&0.05&0.04&0.05&0.05\\

\hline
 \multicolumn{8}{|c|}{IMPROVEMENTS} \\ 
 \hline
 Min.&0.33&0.25&0.25&0.25&0.31&0.25&0.42\\
Max.&0.83&0.81&0.68&0.78&0.76&0.76&0.79\\
Mean&0.64&0.64&0.46&0.58&0.62&0.58&0.65\\
Variance&0.01&0.01&0&0.01&0.01&0.01&0.01\\
Standard deviation&0.09&0.1&0.06&0.08&0.1&0.08&0.07\\

 \hline
 \multicolumn{8}{|c|}{NEW FEATURES} \\ 
 \hline

Min.&0.33&0.33&0.33&0.22&0.33&0.22&0.33\\
Max.&1&0.96&0.67&0.85&0.89&0.89&0.9\\
Mean&0.7&0.69&0.47&0.6&0.68&0.61&0.68\\
Variance&0.01&0.01&0&0.01&0.01&0.01&0.01\\
Standard deviation&0.1&0.09&0.05&0.09&0.08&0.11&0.09\\

 \hline
\end{tabular}
\label{table:AC}
\end{center}
\end{table}

\newpage
\textbf{Precision results of predicting successful and unsuccessful issues}
\begin{table}[hbt!]
\tiny
\begin{center}
 \begin{tabular}{|c|c|c|c|c|c|c|c|}
 \hline
{} & {MNB} & {LR} & {SVC} & {DTC} & {MLPC} & {RFC} & {GBC} \\ 
 \cline{1-8}
 \multicolumn{8}{|c|}{SUCCESSFUL BUGS} \\ 
 \hline
Min.&0.61&0.59&0&0.51&0.59&0.46&0.49\\
Max.&1&1&0.57&0.98&1&1&1\\
Mean&0.77&0.74&0.34&0.72&0.77&0.63&0.69\\
Variance&0.01&0.01&0.02&0.01&0.01&0.02&0.01\\
Standard deviation&0.09&0.1&0.14&0.11&0.1&0.12&0.11\\

 \hline
 \multicolumn{8}{|c|}{UNSUCCESSFUL BUGS} \\ 
 \hline
Min.&0.65&0.71&0.39&0.5&0.64&0.46&0.49\\
Max.&1&0.95&0.57&0.88&0.91&1&1\\
Mean&0.82&0.84&0.45&0.76&0.81&0.63&0.69\\
Variance&0&0&0&0&0&0.02&0.01\\
Standard deviation&0.06&0.05&0.04&0.07&0.05&0.12&0.11\\

 \hline
 \multicolumn{8}{|c|}{SUCCESSFUL IMPROVEMENTS} \\ 
 \hline
Min.&0.08&0.08&-&0.12&0.11&0.05&0\\
Max.&0.86&0.61&-&0.59&0.57&0.51&0.67\\
Mean&0.36&0.31&-&0.27&0.27&0.24&0.3\\
Variance&0.02&0.01&-&0.01&0.01&0.01&0.01\\
Standard deviation&0.14&0.11&-&0.1&0.1&0.1&0.12\\

 \hline
 \multicolumn{8}{|c|}{UNSUCCESSFUL IMPROVEMENTS} \\ 
 \hline

Min.&0.65&0.65&-&0.61&0.54&0.05&0\\
Max.&1&1&-&1&1&0.51&0.67\\
Mean&0.86&0.86&-&0.84&0.85&0.24&0.3\\
Variance&0&0&-&0.01&0&0.01&0.01\\
Standard deviation&0.06&0.06&-&0.07&0.07&0.1&0.12\\

 \hline
 \multicolumn{8}{|c|}{SUCCESSFUL NEW FEATURES} \\ 
 \hline
 Min.&0.03&0.03&-&0&0&0&0\\
Max.&0.62&0.63&-&0.58&0.59&0.54&0.6\\
Mean&0.26&0.26&-&0.22&0.24&0.22&0.25\\
Variance&0.02&0.02&-&0.02&0.02&0.02&0.02\\
Standard deviation&0.15&0.15&-&0.14&0.15&0.12&0.14\\

 \hline
 \multicolumn{8}{|c|}{UNSUCCESSFUL NEW FEATURES} \\ 
 \hline

Min.&0.68&0.7&-&0.62&0.69&0&0\\
Max.&1&1&-&1&1&0.54&0.6\\
Mean&0.89&0.88&-&0.84&0.86&0.22&0.25\\
Variance&0&0&-&0.01&0&0.02&0.02\\
Standard deviation&0.07&0.07&-&0.08&0.07&0.12&0.14\\

 \hline
\end{tabular}
\label{table:PR}
\end{center}
\end{table}

\newpage 
\textbf{Recall results of predicting successful and unsuccessful issues}
\begin{table}[hbt!]
\tiny
\begin{center}
 \begin{tabular}{|c|c|c|c|c|c|c|c|}
 \hline
{} & {MNB} & {LR} & {SVC} & {DTC} & {MLPC} & {RFC} & {GBC} \\ 
 \cline{1-8}
 \multicolumn{8}{|c|}{SUCCESSFUL BUGS} \\ 
 \hline
Min.&0.53&0.66&0&0.46&0.55&0.58&0.64\\
Max.&1&0.95&1&0.87&0.92&0.97&0.96\\
Mean&0.71&0.76&0.86&0.62&0.7&0.7&0.78\\
Variance&0.01&0.01&0.12&0.01&0.01&0.01&0.01\\
Standard deviation&0.11&0.08&0.34&0.07&0.08&0.07&0.08\\

 \hline
 \multicolumn{8}{|c|}{UNSUCCESSFUL BUGS} \\ 
 \hline

Min.&0.76&0.74&0&0.74&0.73&0.61&0.53\\
Max.&1&1&1&0.97&1&1&1\\
Mean&0.86&0.83&0.14&0.84&0.86&0.72&0.77\\
Variance&0&0&0.12&0&0&0.01&0.01\\
Standard deviation&0.05&0.06&0.34&0.06&0.05&0.08&0.08\\

 \hline
 \multicolumn{8}{|c|}{SUCCESSFUL IMPROVEMENTS} \\ 
 \hline

Min.&0.2&0.25&-&0.17&0.25&0.13&0\\
Max.&1&1&-&1&1&1&1\\
Mean&0.52&0.61&-&0.53&0.63&0.58&0.62\\
Variance&0.02&0.02&-&0.02&0.01&0.01&0.02\\
Standard deviation&0.15&0.12&-&0.13&0.11&0.11&0.13\\

 \hline
 \multicolumn{8}{|c|}{UNSUCCESSFUL IMPROVEMENTS} \\ 
 \hline
Min.&0.4&0.33&-&0.23&0.31&0.22&0.42\\
Max.&0.98&0.89&-&0.82&0.65&0.71&0.94\\
Mean&0.73&0.65&-&0.62&0.55&0.53&0.63\\
Variance&0.01&0.01&-&0.01&0&0.01&0.01\\
Standard deviation&0.12&0.1&-&0.09&0.06&0.08&0.09\\

 \hline
 \multicolumn{8}{|c|}{SUCCESSFUL NEW FEATURES} \\ 
 \hline

Min.&0.25&0.25&-&0&0&0&0\\
Max.&1&1&-&1&1&1&1\\
Mean&0.69&0.67&-&0.47&0.58&0.56&0.61\\
Variance&0.02&0.02&-&0.02&0.03&0.02&0.04\\
Standard deviation&0.14&0.14&-&0.16&0.17&0.15&0.2\\

 \hline
 \multicolumn{8}{|c|}{UNSUCCESSFUL NEW FEATURES} \\ 
 \hline

Min.&0.07&0.13&-&0.18&0.25&0.17&0.19\\
Max.&0.72&0.7&-&0.92&0.74&0.69&0.84\\
Mean&0.53&0.54&-&0.6&0.55&0.52&0.59\\
Variance&0.03&0.02&-&0.02&0.01&0.01&0.01\\
Standard deviation&0.16&0.15&-&0.13&0.11&0.09&0.11\\

 \hline
\end{tabular}
\label{table:RE}
\end{center}
\end{table}

\newpage
\textbf{F1-score results of predicting successful and unsuccessful issues}
\begin{table}[hbt!]
\tiny
\begin{center}
 \begin{tabular}{|c|c|c|c|c|c|c|c|}
 \hline
{} & {MNB} & {LR} & {SVC} & {DTC} & {MLPC} & {RFC} & {GBC} \\ 
 \cline{1-8}
 \multicolumn{8}{|c|}{SUCCESSFUL BUGS} \\ 
 \hline
Min.&0.6&0.65&0&0.53&0.62&0.54&0.59\\
Max.&1&0.96&0.72&0.92&0.94&0.93&0.96\\
Mean&0.73&0.75&0.48&0.66&0.73&0.66&0.73\\
Variance&0.01&0.01&0.04&0.01&0.01&0.01&0.01\\
Standard deviation&0.1&0.08&0.2&0.08&0.08&0.09&0.09\\

 \hline
 \multicolumn{8}{|c|}{UNSUCCESSFUL BUGS} \\ 
 \hline

Min.&0.73&0.73&0.57&0.6&0.69&0.61&0.64\\
Max.&1&0.96&0.72&0.91&0.94&0.91&0.95\\
Mean&0.84&0.83&0.62&0.79&0.83&0.75&0.8\\
Variance&0&0&0&0&0&0&0\\
Standard deviation&0.05&0.04&0.04&0.05&0.04&0.05&0.06\\

 \hline
 \multicolumn{8}{|c|}{SUCCESSFUL IMPROVEMENTS} \\ 
 \hline
Min.&0.13&0.13&-&0.15&0.18&0.07&0\\
Max.&0.63&0.64&-&0.57&0.63&0.6&0.67\\
Mean&0.4&0.4&-&0.35&0.37&0.33&0.39\\
Variance&0.01&0.01&-&0.01&0.01&0.01&0.01\\
Standard deviation&0.1&0.1&-&0.1&0.11&0.1&0.11\\

 \hline
 \multicolumn{8}{|c|}{UNSUCCESSFUL IMPROVEMENTS} \\ 
 \hline

Min.&0.55&0.45&-&0.38&0.44&0.3&0.57\\
Max.&0.93&0.89&-&0.83&0.76&0.77&0.91\\
Mean&0.78&0.73&-&0.71&0.67&0.64&0.72\\
Variance&0.01&0.01&-&0&0&0.01&0\\
Standard deviation&0.07&0.07&-&0.07&0.05&0.08&0.06\\

 \hline
 \multicolumn{8}{|c|}{SUCCESSFUL NEW FEATURES} \\ 
 \hline
 
 Min.&0.05&0.05&-&0&0&0&0\\
Max.&0.64&0.65&-&0.56&0.64&0.58&0.64\\
Mean&0.35&0.35&-&0.28&0.32&0.3&0.33\\
Variance&0.02&0.02&-&0.02&0.02&0.02&0.02\\
Standard deviation&0.14&0.14&-&0.13&0.15&0.13&0.14\\

 \hline
 \multicolumn{8}{|c|}{UNSUCCESSFUL NEW FEATURES} \\ 
 \hline
Min.&0.13&0.22&-&0.29&0.38&0.28&0.33\\
Max.&0.79&0.78&-&0.86&0.79&0.77&0.9\\
Mean&0.64&0.66&-&0.69&0.66&0.64&0.69\\
Variance&0.02&0.02&-&0.01&0.01&0.01&0.01\\
Standard deviation&0.15&0.13&-&0.1&0.08&0.08&0.09\\

 \hline
\end{tabular}
\label{table:F1C}
\end{center}
\end{table}

\newpage

  \bibliographystyle{elsarticle-harv} 
  \bibliography{Main_text}





\end{document}